RESEARCH ARTICLE / ARAŞTIRMA MAKALESİ

# Phillips Curve Estimation During Tranquil and Recessionary Periods: Evidence From Panel Analysis*

## Sakin Dönem ve Durgunluk Dönemi Phillips Eğrisi Tahmini: Panel Analizinden Kanıtlar

Yhlas SOVBETOV[1]

**ABSTRACT**

The empirical literature that covers Phillips Curve analysis during recessionary periods is notably scant. The Great Recession has rekindled a debate on the validity and stability of the Phillips Curve which is still ongoing. The basis for this debate is the observation that real activity dropped sharply without causing a drop in inflation. This paper carries out an empirical analysis for the classical expectation-augmented Phillips curve model across 41 countries from 1980-2016 by distinguishing tranquil and recessionary periods separately. Based on the results of the research, the paper finds that dynamics of Phillips Curve changes during recessionary periods and the empirical relationship becomes no longer valid. These findings support the ongoing debate about the missing disinflation and collapse of the Phillips curve, but only during the recessionary periods. In the case of tranquil periods, the empirical relationship still seems to be valid. Moreover, the paper also observes that both backward-looking and forward-looking fractions of inflation gain weight and significance during recessionary periods. However, the paper remains indecisive about which exact fraction gains more weight and significance as the panel model does not incorporate these two fractions of inflation in a single hybrid framework simultaneously.

**Keywords:** Phillips curve, Panel analysis, Unemployment
**JEL Classification:** C10, E10, E19

**ÖZ**

Resesyon dönemlerinde Phillips Eğrisi analizini kapsayan ampirik literatür oldukça azdır. Büyük Durgunluk Phillips Eğrisi'nin geçerliliği ve istikrarı üzerine halen devam etmekte olan bir tartışmayı yeniden alevlendirmiştir. Bu tartışmanın temeli ise, gerçek iktisadi faaliyetlerin keskin bir şekilde düşerken enflasyonda bir düşüşün olmamasının gözlemlenmesidir. Bu makale, sakin ve durgunluk dönemlerini ayırarak 1980'den 2016 yılına kadar 41 ülke üzerinde, beklentilerle genişletilmiş klasik Phillips Eğrisi modeli için ampirik çalışma yürütmektedir. Araştırmanın sonuçlarına dayanarak, bu makale, Phillips Eğrisi dinamiğinin durgunluk dönemlerinde değiştiğini ve ampirik ilişkinin artık geçerli olmadığını ortaya koymaktadır. Bu sonuçlar, gözlemlenemeyen enflasyon düşüşü ve Phillips eğrisinin çöküşü konusundaki süregelen tartışmaları sadece durgunluk dönemleri için desteklemektedir. Sakin iktisadi dönemlerdeki durumda ise, bu ampirik ilişkinin hala geçerli olduğu görülmektedir.

Ayrıca, bu makale hem geriye dönük hem de ileriye dönük enflasyon bileşeninin durgunluk dönemlerinde ağırlık ve önem kazandığını da gözlemlemektedir. Bununla birlikte, panel modeli enflasyonun bu iki bileşenini aynı anda tek bir melez çerçeveye dâhil etmediği için bu araştırma hangi bileşenin daha fazla ağırlık ve önem kazandığı konusunda kesin bir sonuç verememektedir.

**Anahtar kelimeler:** Phillips eğrisi, Panel analizi, İşsizlik
**JEL Sınıflaması:** C10, E10, E19

## 1. Introduction

The strong impact of business cycles on inflation and unemployment is a known phenomenon. Sharp changes in these macroeconomic indicators during recent global slowdowns are clear examples for business cycle impact. High unemployment also indicates an inefficient use of resources which causes negative output gaps and price levels to drop.

This interaction between unemployment rate and price levelshas been a matter of interest to many policy-makers for many years ever since its discovery by William Phillips (1958) who observed an inverse correlation: when unemployment goes up, wages start to slowly decrease, and when unemployment drops to low levels, wages tend to rapidly rise. He believes that this happens due to the looseness of the labor market when unemployment rates are high, and tightness when rates are lower. Thereafter, this empirical finding has become known as the Phillips Curve and has become a fixture in many macroeconomics textbooks.

Although the Phillips Curve (PC) was widely accepted and used by policy makers who wanted to benefit from the empirical trade-off in the early 1960s, it has received a lot of criticism. Indeed, these comments helped the originally portrayed model to evolve over time reflecting the theoretical developments of the last half-century. For instance, Phelps (1967) and Friedman (1968) criticize the PC arguing that the presented trade-off might occur only in the short run. Moreover, equilibrium in the labor market is determined by real wage, so PC won't work as it accounts only for money wages (Akerlof, 2007). Moreover, Phelps and Friedman believe that structural and frictional unemployment is never-ending, so it is the *"natural rate"* of unemployment at which inflation rate remains stable (later it became known as the non-accelerating inflation rate of unemployment, NAIRU). If government persistently generates inflationary policies to reduce unemployment below this natural rate, there will be a short-run trade-off (Samuelson and Solow, 1960) which is today known as the *"Expectation-Augmented Phillips Curve"* (EAPC). However, after a time, individuals start to show evidence of *adaptive expectations (backward-looking)* by adjusting their current expectations based on past years' inflation. Thus, in turn, it reverts unemployment rate back to its natural level. Therefore, the curve is vertical in the long-run, thus unemployment is irrelevant to the level of inflation.

Later, Lucas (1976) posits a critique ignoring the short-run trade-off of PC arguing that agents have *rational expectations (forward-looking)* rather than adaptive ones. Therefore, they account not only for recent information, but all available information, and adjust their expectation instantaneously, so that short-run trade-off between inflation and unemployment does not occur.

The validity and stability of this empirical relationship is still debated nowadays. The majority of recent literature (Russell and Banerjee, 2008; Paul, 2009; Stock, 2011; IMF, 2013; Ojapinwa and Esan, 2013; Watson, 2014; Yellen, 2015; Kiley, 2015; Krugman, 2015; Coibion and Gorodnichenko, 2015; Blanchard, 2016; Mazumder, 2018; Murphy, 2018; Ball and Mazumder, 2019; Sovbetov and Kaplan, 2019a, 2019b) reports a time and cross-section variability in the Phillips Curve. However, reasons for this variability have not been clearly addressed yet. This paper aims to examine Phillips relationship over a panel of large sampled countries placing a special focus on tranquil and recessionary periods. It is due to the fact that a changing regime or environment might be the main reason for variability in the Phillips Curve.

The rest of the paper is structured in the following order. The next section briefly reviews related literature. The third section describes data and specifies the methodology for this study. The fourth section presents the findings and interprets them thoroughly, and the final section concludes.

**2. Literature Review**

Although much of recent literature finds that inflation dynamics are forward-looking, and Phillips curve works as the theory suggests, the remaining portion of literature documents empirical results that cast doubt on the validity of the Phillips curve. For instance, Fendel et al. (2011) test both the traditional and expectation-augmented version of Phillips Curve in G7 countries between October 1989 and December 2007. They observe a significant trade-off between inflation and unemployment throughout all the G7 countries, except Italy where the traditional Phillips Curve fails to work. However, when they pursue the same analysis with the expectation-augmented Phillips Curve they find out that it properly works for all G7 countries, with the strongest trade-off appearing in Japan and in the US. Rulke (2012) also finds a similar trade-off with the expectation-augmented Phillips Curve throughout six Asian-Pacific countries. He observes that the trade-off magnitude is remarkably larger in Japan and South Korea.

Russell and Banerjee (2008) study the expectation-augmented Phillips curve under non-stationarity conditions in the series. They observe a positive relationship between inflation and unemployment rate in the short run for the United States, so they conclude that the Phillips Curve does not work. Paul (2009) also fails to document the existence of an empirical Phillips curve in India. He states that the relationship is often evasive or absent in less-developed economies. Similar conclusions were made by Sovbetov and Kaplan (2019b) who have studied the Phillips curve in 41 different countries during tranquil and recessionary periods. They note that the Phillips curve might not work as notionally in less-developed or crisis-prone countries due to a lack of well-established and freely operating structure of macroeconomic foundations and motivations. On the other hand, Ojapinwa and Esan (2013) find a weak Phillips Curve trade-off for Nigeria in the short run, but it disappears in the long run as inflation and unemployment move together positively.

During the Great Recession several economists argued that the Phillips Curve relationship seemed to have broken down. The basis for this argument is the observation that real activity dropped sharply without causing a drop in inflation. They also state that the Phillips curve failed to explain the missing disinflation due to anchored expectations (Stock, 2011; International Monetary Fund [IMF], 2013; Watson, 2014; Yellen, 2015; Kiley, 2015; Krugman, 2015; Blanchard, 2016; Mazumder, 2018; Ball and Mazumder, 2019). For instance, Ball and Mazumder (2011) argue that Phillips curves estimated over the period 1960-2007 in the US cannot explain the behavior of inflation in the period 2008-2010. Moreover, they conclude that the Great Recession provides fresh evidence for the instability of the Phillips curve. They argue the fact that the fit of that Phillips equation deteriorates once data for the years 2008-2010 are added to the sample. Similarly, Bulligan and Viviano (2017) examine whether the Phillips relationship has changed in Euro Area since the Great Recession of 2008. They find evidence that the wage Phillips curve has changed since the great financial crisis and the correlation between wage inflation and unemployment rate has increased in Italy, France and Spain while such correlation has diminished in Germany.

On the other hand, Del Negro et al (2015) challenges this argument by showing that this observation can be reconciled with predictions of the Phillips curve model. They propose a new model and argue that marginal costs will revert back to a steady state after the crisis, which, through the forward-looking Phillips curve, prevents a prolonged deflationary episode. However, Van Zandweghe (2019) underlines that Del Negro et al.'s model predicts a stable unit labor cost, which had declined during the Great Recession.

Conti and Gigante (2018) investigate the dynamics of core inflation in Italy between 1999Q1 and 2017Q1 periods with a special focus on periods after the Great Recession. As a result, they document significant trade-off between core inflation and economic activity when labour market slacks are fixed. They also observe a steepening in the Phillips curve after the Global Financial Crisis, but when checking for financial indicators, the slope of the Phillips curve turns out to be flatter. Thus, they conclude that financial indicators help to better characterize the dynamics of core inflation.

To conclude, behavioural changes in the dynamics of inflation, particularly in recessionary periods, have been growing topics in Phillips curve literature since the recent global recession that led to large and persistent output gaps. In this context, examining the validity and stability of the Phillips curve by distinguishing tranquil and recessionary periods over a large country sample might make a crucial contribution to the relevant field of literature.

### 3. Data and Methodology

The paper empirically tests the validity and stability of backward-looking and forward-looking Phillips curves between Q1:1980 and Q1:2016 across 41 countries (a list of sampled countries are given in Appendix Table A1) by placing a particular focus on tranquil and recessionary periods. It is important in order to check whether the empirical relationship behaves notionally or changes during recessionary periods.

The word *"recession"* in this study refers to all non-growing economic periods, while the remaining periods are defined as *"tranquil"* periods. We follow this basic approach because during economic recessions various aspects of the economy are disrupted, so the study aims to capture all their influences over the Phillips relationship through changed expectations.

The research carries this analysis in a panel framework by controlling country-specific effects. It is reasonable to assume that developed, emerging, and frontier markets naturally have different idiosyncratic characteristics. So, fixing all these country-specific effects in a panel analysis should derive more accurate results compared to a generalization of country-based OLS results.

In order to specify the panel model, we examine consistency and efficiency of GLS estimators through cross-section fixed (FE) and random effects (RE) as well as pooled estimation. All these specifications have potential advantages and disadvantages (see Kinney and Dunson, 2007; Park, 2009; Bondell et al., 2011; Clark and Linzer, 2015; Bell and Jones, 2015). The pooled model often generates biased estimates as it does not impose any controls for between-effects (unobserved heterogeneity between cross-sections such as culture, religion, gender, race, etc.) among cross-sections which leads its residuals to be heteroskedastic. The cross-section fixed-effects model assumes that heterogeneity among all entities is time-invariant and fixed across entities, and it wipes out these between-effects (unobserved heterogeneity among cross-section) by allowing each entity to have their own intercept values. In other words, the fixed dummy variable controls for all time-invariant differences between entities where the intercept varies among cross-sections and remains stable over the time horizon. In this respect, the FE model specifications for backward- and forward-looking Phillips Curve models can be formed by panelising the Phillips curve model and adding a cross-section fixed effects dummy as follows.

$$\pi_{it} = \beta_0 + \beta_1 \pi_{it}^e + \beta_2 U_{GAP\ it} + \beta_3 \pi_{it}^e DUMMY_R + \beta_4 U_{GAP\ it} DUMMY_R + \alpha_i D_i + \varepsilon_{it} \quad (1)$$

where *DUMMY* is a dummy variable that is attributed a value of 1 during recessionary periods and zero during other periods; $\pi_t$ and $\pi_t^e$ are proxied by first differences of logarithm of *CPI* and expected *CPI inflation* over one year respectively; $U_t$, and $U^*$ are proxied by unemployment rate and *NAIRU* respectively in logarithmic form. It is worth noting that NAIRU series are derived by HP filtering[1] the current inflation (first column) with lambda 1600. This is a common methodology to strip out trend components from the cyclical one. Data for these variables are obtained from Thomson Reuters Eikon

---

[1] HP is a technique to derive long-run levels of variables. The $\lambda$ is a smoothing parameter that is set by using Ravn and Uhliq (2002) frequency rule: the number of periods per year divided by 4, raised to the power *x*, and multiplied by 1600. Hodrick and Prescott (1997) recommend the value 2 for *x*, whereas Ravn and Uhliq (2002) suggest using 4 for *x*. Following Hodrick and Prescott (1997), we derive *λ=1600* for our dataset.

DataStream, and a fixed constant term is added to all series to handle negative values during transformation into logarithmic form which only shifts $\beta_0$ up leaving other variables unaffected. Additionally, we show stationarity of all related series under Augmented Dickey-Fuller (ADF) and Phillips-Perron (PP) tests in appendix Table A2.

The $\beta_2$ and $\beta_2+\beta_4$ indicate Phillips coefficients during tranquil and recessionary periods respectively. Similarly, $\beta_1$ and $\beta_1+\beta_3$ show fractions of $\pi^e$ in current inflation during tranquil and recessionary periods respectively. Note that if $\pi_t^e$ equates to $\pi_{t-1}$, then the model converts to a Phillips Curve with backward-looking specification. And if it equates to expected inflation $E_t(\pi_{t+1})$, then the model takes the shape of Phillips Curve with forward-looking specification. In addition, the study uses manual calculations of standard errors $\beta_2+\beta_4$ and $\beta_1+\beta_3$ with formulas of $\sigma_{\beta2+\beta4} = \sqrt{\sigma_{\beta2}^2 + \sigma_{\beta4}^2}$ and $\sigma_{\beta1+\beta3} = \sqrt{\sigma_{\beta1}^2 + \sigma_{\beta3}^2}^2$ in order to calculate the significance of the Phillips coefficients during recessionary periods.

The $D_i$ is a dummy variable which takes 1 for the country "$i$" ($i=1,...41$), and zero for other countries in the sample. Moreover, if our panel data was unbalanced, we would also include a fixed effect for the period by considering another dummy variable to proxy years as "$D_y$". The $\beta_0+\alpha_i D_i$ controls time-invariant between-effect (cross-section) and $\varepsilon_{it}$ is an idiosyncratic error term. Also, note that $\pi_t^e$ equates to $\pi_{t-1}$ in the backward-looking model, and it equates to expected inflation $E_t(\pi_{t+1})$ in the forward-looking model.

The FE specification is based on two main assumptions: (1) $\epsilon t \sim i.i.d.N(0,\sigma_e^2)$, and (2) exogeneity of covariates $x_{ij}$, i.e. $cov(\epsilon_{ij}, x_{ij})=0$ for all dependent variables where $Var(\epsilon_{ij})=\sigma_e^2$. However, the residuals of the FE model might not always be the most efficient (although it is consistent) as they completely wipe out between-effects and their residuals account only for within-effects (heterogeneity within cross-section, i.e. among variables of same cross-section).

The random-effect model is a specific type of pooled estimation that assumes time-varying heterogeneity across entities, thus, it generates estimates closer, on average, to the true value of any particular country. In this respect, we establish our RE specified model by modifying the Eq. (1) equation as below.

$$\pi_{it} = \beta_{0,i} + \beta_1 \pi_{it}^e + \beta_2 U_{GAP\ it} + \beta_3 \pi_{it}^e DUMMY_R + \beta_4 U_{GAP\ it} DUMMY_R + \varepsilon_{it} \qquad (2)$$

where $\beta_{0,i} = \beta_0 + \omega_i$ with $\omega_i \sim N(0, \sigma^2)$. And when $\beta_{0,i}$ is plugged into the above equation, it becomes as follows:

$$\pi_{it} = \beta_0 + \beta_1 \pi_{it}^e + \beta_2 U_{GAP\ it} + \beta_3 \pi_{it}^e DUMMY_R + \beta_4 U_{GAP\ it} DUMMY_R + u_{it} \qquad (3)$$

where $u_{i,t} = \omega_i + \varepsilon_{i,t}$. The $\omega_i$ controls between-entity errors, while $\varepsilon_t$ controls within-entity errors. The residuals of the RE model are often smaller, and thus homoscedastic, and it also offers the possibility of a differential between cross-sections. However, the estimators of this RE specification are often biased due to potential correlation between covariates of explanatory variables and $\omega_i$. Unlike the FE model, it captures both *"within"* and *"between"* deviations, and allows all cross-sections to have a common mean value for the intercept. In other words, the dummy variable *"$D_i$"*- which was a part of the intercept in the FE - becomes a part of the error *"$u_i$"* in the RE model.

A priori researchers' preference in the trade-off between bias and variance, it is more logical to exhibit the dataset and characteristics of the sample. Additionally, there are some statistical tests that might be a guideline in the selection of an appropriate model (see Table 1). Following one of these guidelines, we initially employ Redundant Fixed Effect and Breush-Pagan LM tests to find out whether our panel data contain a fixed effect and a random effect respectively. In a special case, when both fixed and random effects are observed, we schedule the Hausman (1978) test for the final decision-making in model selection as below.

$$H = (\beta_1 - \beta_0)'(Var(\beta_1) - Var(\beta_0))^p (\beta_1 - \beta_0) \qquad (4)$$

where $\rho$ is pseudo-inverse. Hausman test hypothesizes the null hypothesis to alternative where $H_0$ specifies that both FE ($\beta_0$) and RE estimators ($\beta_1$) are consistent, but RE is more efficient than FE, while the alternative hypothesis specifies that RE is inconsistent, and only FE ($\beta_0$) is consistent.

Table 1: Fixed and Random Effects Model Selection

| Redundant Fixed Effect Test | Breusch-Pagan & Honda LM Tests | Concluded Model |
| --- | --- | --- |
| $H_0$ is not rejected (No fixed effect) | $H_0$ is not rejected (No random effect) | Data are poolable (Pooled OLS) |
| $H_0$ is rejected (Fixed effect) | $H_0$ is not rejected (No random effect) | Fixed Effect Model (GLS) |
| $H_0$ is not rejected (No fixed effect) | $H_0$ is rejected (Random effect) | Random Effect Model (GLS) |
| $H_0$ is rejected (Fixed effect) | $H_0$ is rejected (Random effect) | (1) Both Fixed and Random Effect Models (2) Hausman Test (recommended) |

**Notes:** The null hypothesis for both Breush - Pagan and Honda LM tests are "No Random Effects". The null hypothesis for Redundant Fixed Effect test is "No Unobserved Heterogeneity (No Fixed Effect)".

## 4. Results and Discussion

Following the guidelines in Table 1, we ran Redundant Fixed Effect and Breush-Pagan LM tests for both backward- and forward-looking Phillips Curves. As an after effect, we found that both tests signaled the existence of fixed- and random-effects at a 1% significance level. Therefore, we conducted the Hausman test for final decision making in model selection. We reported the results of the Hausman test given at Eq.(4) in Table 2 where the outcome recommends using the FE specification for both backward- and forward-looking Phillips Curve models by rejecting the null hypothesis with a probability close to zero.

Table 2: Results for Correlated Random Effects - Hausman Test

| Model | Redundant Fixed Effect | Breush-Pagan LM | Hausman Test |
| --- | --- | --- | --- |
| BL | 187,8254 (0,0000) | 182,0554 (0,0000) | 209,3493 (0,0000) |
| FL | 215,0347 (0,0000) | 178,9009 (0,0000) | 87,7315 (0,0000) |

**Notes:** BL and FL represent Backward-Looking and Forward-Looking models respectively. The Chi-square statistics are given in the columns with probabilities of rejecting null hypothesis in parentheses. The null hypothesis of the Redundant Fixed Effect test is "no fixed effects"; whereas the null hypothesis of Breush-Pagan LM test is "no random effects". The null hypothesis of the Hausman test is $H_0$: Both FE and RE estimators are consistent, but RE estimators are more efficient than FE. The alternative hypothesis of the Hausman test is $H_1$: RE estimators are inconsistent, FE estimators are consistent. The Hausman degree of freedom of chi-square for both models is 4.

Yhlas SOVBETOVEven though Hausman test suggests using the FE specification, we reported the results of pooled, RE, and FE panel OLS estimations of Phillips Curve for the whole sample (41 countries) in order to make a broad comparison in Table 3. Notice that all panel models (pooled, RE, and FE) generate close estimates for inflation dynamics, but the appropriate one is the FE specification in both backward- and forward-looking cases. The R-square value of FE backward-looking specification indicates that explanatory factors account for nearly 68% of variations in current inflation, and the Phillips coefficient is estimated as -0,07 during growing economic periods at the 5% significance level. It indicates that a 1%increase in unemployment gap decreases inflation by 0,07%. The Phillips coefficient, however, sharply changes to -0,14 and totally loses its significance during recessionary periods. The coefficient of $\pi^e_t$ during non-recessionary periods takes a value of 0,4759 at the 1% significance level which indicates that 47,59% of current inflation ($\pi_t$) is formed by previous period's inflation ($\pi_{t-1}$). This increases to 0,8441 during recessionary periods preserving its significance at 1% level. This implies that the share of past inflation ($\pi_{t-1}$) in current inflation ($\pi_t$) jumps from 47,59% to 84,41%. The pooled and RE specifications also derive similar results due to the Swamy-Arora (1972) estimator of the variance components of RE where $\sigma_u$ (cross-section random) takes a rho number of zero and $\sigma_e$ (idiosyncratic random) takes1. This means the variance of RE is entirely comprised of idiosyncratic random, and the weight of cross-section random is zero. Thus, both pooled and RE models generate similar results.

Table 3: OLS Panel Estimation of EAPC during Normal/Recession Periods

| Variables | Backward-looking | | | Forward-looking | | |
|---|---|---|---|---|---|---|
| | Pooled | Fixed | Random | Pooled | Fixed | Random |
| Intercept | 0,0032*** (0,0007) | 0,0037*** (0,0007) | 0,0032*** (0,0007) | 0,0030*** (0,0002) | 0,0033*** (0,0005) | 0,0031*** (0,0005) |
| $^N\pi^e_t$ | 0,5369*** (0,0750) | 0,4759*** (0,0679) | 0,5369*** (0,0750) | 0,5503*** (0,0110) | 0,5097*** (0,0778) | 0,5373*** (0,0777) |
| $^NU_{GAP}$ | -0,0899*** (0,0372) | -0,0714** (0,0327) | -0,0899*** (0,0372) | -0,0532** (0,0236) | -0,0380** (0,0191) | -0,0487** (0,0211) |
| $^R\pi^e_t$ | 0,8848*** (0,1326) | 0,8441*** (0,1227) | 0,8848*** (0,1326) | 0,6281*** (0,0162) | 0,6142*** (0,1129) | 0,6239*** (0,1149) |
| $^RU_{GAP}$ | -0,1507 (0,1150) | -0,1411 (0,1148) | -0,1507 (0,1150) | -0,0929 (0,0554) | -0,0829 (0,0927) | -0,0902 (0,0930) |
| Weighted $R^2$ | - | - | 0,6676 | - | - | 0,7433 |

| Unweighted R² | 0,6676 | 0,6802 | 0,6676 | 0,7560 | 0,7641 | 0,7559 |
| --- | --- | --- | --- | --- | --- | --- |
| Panel Obs. | 5587 | 5587 | 5587 | 5591 | 5591 | 5591 |
| $\sigma_u$ | - | - | 0,0000 | - | - | 0,0043 |
| $\sigma_e$ | - | - | 1,0000 | - | - | 0,9957 |

**Notes:** The Panel OLS estimation methodology is used to determine dynamics of inflation, $\pi_t$, with 41 cross-sections and 144 periods (1980Q2-2016Q1). The $\pi^e_t$ equates to past inflation ($\pi_{t-1}$) in backward-looking model, while it takes value of $E_t(\pi_{t+1})$ in the forward-looking case. The superscripts "N" and "R" indicate estimates for normal (growing) and recessionary periods respective. The numbers in the table are estimated coefficients with white standard errors and covariance in parenthesis, and *, **, and *** denote significance at 10%, 5%, and 1% levels respectively. The $\sigma_u$ and $\sigma_e$ represent the Swamy and Arora estimator of variance components of random effect (cross-section and idiosyncratic respectively) with rho numbers.

On the right hand side of Table 3, the R-square value of FE forward-looking specification account for the relatively higher (76%) variations of current inflation compared to the backward-looking model. The Phillips coefficient shrinks to nearly half; it is about -0,038 significant at 5% level. Similar to the backward-looking case, the coefficient doubles in absolute magnitude (-0,0829) and loses its statistical significance during recessionary periods. The Phillips coefficient indicates that a percentage increase in unemployment gap decreases inflation by 0,04%.

The coefficient of $\pi^e_t$ during growing periods takes a value of 0,5097 at 1% significance level which indicates that 51% of current inflation ($\pi_t$) is formed by expected future inflation ($E_t(\pi_{t+1})$). This coefficient increases to 0,6142 during non-growing periods preserving its significance at 1% level. A similar scenario is observed in the backward-looking case. This indicates that the share of expected future inflation ($E_t(\pi_{t+1})$) in current inflation ($\pi_t$) jumps from 50,97% to 61,42%. It indicates that inflation becomes more sensitive to expected inflation (to lagged inflation in backward-looking case) and Phillips relation collapses during recessionary periods. The pooled and RE specifications also derive alike results as $\sigma_u$ (cross-section random) takes rho number of zero and $\sigma_e$ (idiosyncratic random) takes rho number of 1.

The panel results reveal that the Phillips relation collapses during recessionary periods and inflation becomes more sensitive to its lagged (in the case of backward-looking) and expected future values (in the case of forward-looking). Although absolute magnitudes of $\pi_{t-1}$ and $E_t(\pi_{t+1})$ increase about 0,36 and 0,11 respectively during recessionary periods, it is unclear which fraction gains more weight and significance. In other words, the results do not imply that inflation becomes more backward-looking during these periods as models do not incorporate these two variables in a hybrid framework.

## 5. Conclusion

This research examines the validity and stability of the Phillips Curve during tranquil and recessionary periods under a panel of 41 different countries with developed, emerging, and frontier markets. Based on the results of this research, we find that the dynamics of both backward-looking and forward-looking specifications of the Phillips Curve change during recessionary periods and the empirical relationship is no longer valid. This result is in line with the findings of Mazumder (2018), Ball and Mazumder (2019) and Sovbetov and Kaplan (2019b).

In particular, we observe that both past and future expected inflation components gain weight and significance, while the unemployment component loses weight. However, we cannot conclude which fraction - backward-looking or forward-looking - gains more importance during recessionary periods as our panel model does not incorporate these two fractions of inflation in a single hybrid framework simultaneously. Therefore, this creates an opportunity for future research examining the Phillips curve during recessionary periods under a hybrid New Keynesian framework.

To conclude, the evidence documented in this research can be another example for business cycle impact on unemployment, and so, on inflation. This has two implications for monetary policy makers. First, the empirical inflation-unemployment trade-off remains a useful tool for central bankers only during tranquil periods. Second, the evidence suggests that the trade-off disappears during recessionary periods and backward- and forward-looking fractions of inflation become more credible.

**Grant Support:** The author received no financial support for this work.

Watson, M. W. (2014). Inflation Persistence, the NAIRU, and the Great Recession. *American Economic Review*, *104*(5), 31–36. Retrieved from: https://doi.org/10.1257/aer.104.5.31

Yellen, J. L. (2015). Inflation dynamics and monetary policy: A speech at the Philip Gamble Memorial Lecture. University of Massachusetts, Amherst, Massachusetts. *Speech 863*, Board of Governors of the Federal Reserve System (U.S.).

## APPENDICES

### Table A1: Country Codes and Number of Recessions

| Country Name | Code | 1980-2016 |
|---|---|---|
| | | 145 quarters |
| Argentina | AG | 55 |
| Australia | AU | 15 |
| Germany | BD | 42 |
| Belgium | BG | 23 |
| Brazil | BR | 51 |
| Canada | CH | 11 |
| Chile | CL | 32 |
| China | CN | 23 |
| Czech Republic | CZ | 24 |
| Denmark | DK | 50 |
| Spain | ES | 32 |
| Finland | FN | 43 |
| France | FR | 22 |
| Greece | GR | 75 |
| Hungary | HN | 27 |
| Indonesia | ID | 17 |
| India | IN | 24 |
| Ireland | IR | 47 |
| Italy | IT | 47 |
| Japan | JP | 46 |
| South Korea | KO | 11 |
| Mexico | MX | 31 |
| Malaysia | MY | 13 |
| Netherlands | NL | 31 |
| Norway | NW | 44 |
| Austria | OE | 32 |
| Philippines | PH | 21 |
| Poland | PO | 15 |
| Portugal | PT | 37 |
| Romania | RM | 56 |
| Russia | RS | 37 |
| South Africa | SA | 28 |
| Sweden | SD | 32 |

| Singapore | SP | 24 |
|---|---|---|
| Switzerland | SW | 31 |
| Thailand | TH | 29 |
| Turkey | TK | 47 |
| Taiwan | TW | 30 |
| United Kingdom | UK | 18 |
| United States | US | 18 |
| Venezuela | VE | 53 |

**Notes:** Numbers in the table show the quarter numbers with negative GDP growth (recession). The "-" denote missing data.

### Table A2: Results of Unit Root Tests

|  | ADF (intercept) | | | PP (intercept) | | |
|---|---|---|---|---|---|---|
|  | CPI | EI | U_U' | CPI | EI | U_U' |
| AG | 0,0791 (L:2\|N:126) | 0,0508 (L:2\|N:126) | 0,0001 (L:0\|N:144) | 0,0000 (B:7\|N:128) | 0,0000 (B:7\|N:128) | 0,0000 (B:2\|N:144) |
| AU | 0,0008 (L:1\|N:143) | 0,0000 (L:0\|N:144) | 0,0010 (L:2\|N:142) | 0,0000 (B:8\|N:144) | 0,0000 (B:7\|N:144) | 0,0423 (B:6\|N:144) |
| BD | 0,1003 (L:3\|N:141) | 0,0355 (L:3\|N:141) | 0,0419 (L:4\|N:140) | 0,0000 (B:10\|N:144) | 0,0000 (B:10\|N:144) | 0,0008 (B:9\|N:144) |
| BG | 0,0000 (L:0\|N:144) | 0,0350 (L:1\|N:143) | 0,4779 (L:3\|N:141) | 0,0000 (B:9\|N:144) | 0,0001 (B:9\|N:144) | 0,0718 (B:3\|N:144) |
| BR | 0,2191 (L:2\|N:103) | 0,115 (L:3\|N:103) | 0,0000 (L:8\|N:136) | 0,0993 (B:1\|N:105) | 0,2073 (B:6\|N:106) | 0,0000 (B:8\|N:144) |
| CH | 0,0045 (L:4\|N:140) | 0,0040 (L:4\|N:140) | 0,0000 (L:1\|N:143) | 0,0000 (B:10\|N:144) | 0,0000 (B:10\|N:144) | 0,0172 (B:5\|N:144) |
| CL | 0,7175 (L:7\|N:137) | 0,7916 (L:7\|N:137) | 0,0002 (L:1\|N:119) | 0,0000 (B:8\|N:144) | 0,0001 (B:7\|N:144) | 0,0007 (B:4\|N:120) |
| CN | 0,0173 (L:3\|N:141) | 0,0106 (L:2\|N:142) | 0,0431 (L:0\|N:144) | 0,0000 (B:8\|N:144) | 0,0084 (B:4\|N:144) | 0,0241 (B:4\|N:144) |
| CZ | 0,0677 (L:3\|N:96) | 0,0002 (L:0\|N:100) | 0,0081 (L:5\|N:87) | 0,0000 (B:3\|N:99) | 0,0000 (B:17\|N:100) | 0,0509 (B:3\|N:92) |
| DK | 0,0361 (L:4\|N:140) | 0,0273 (L:3\|N:141) | 0,5460 (L:1\|N:143) | 0,0000 (B:10\|N:144) | 0,0000 (B:9\|N:144) | 0,2476 (B:7\|N:144) |
| ES | 0,3683 (L:7\|N:137) | 0,0393 (L:7\|N:137) | 0,6306 (L:1\|N:143) | 0,0000 (B:8\|N:144) | 0,0003 (B:10\|N:144) | 0,3053 (B:7\|N:144) |
| FN | 0,0154 (L:4\|N:140) | 0,0120 (L:4\|N:140) | 0,0001 (L:4\|N:140) | 0,0000 (B:10\|N:144) | 0,0001 (B:10\|N:144) | 0,0282 (B:9\|N:144) |
| FR | 0,0308 (L:11\|N:133) | 0,0239 (L:0\|N:144) | 0,0159 (L:1\|N:143) | 0,0055 (B:9\|N:144) | 0,0298 (B:12\|N:144) | 0,0371 (B:4\|N:144) |
| GR | 0,3312 (L:4\|N:140) | 0,5571 (L:4\|N:140) | 0,0000 (L:8\|N:136) | 0,0000 (B:10\|N:144) | 0,0000 (B:11\|N:144) | 0,036 (B:7\|N:144) |
| HN | 0,3649 (L:3\|N:141) | 0,4055 (L:6\|N:138) | 0,0098 (L:1\|N:99) | 0,0000 (B:10\|N:144) | 0,0000 (B:11\|N:144) | 0,0714 (B:0\|N:100) |
| ID | 0,0000 (L:0\|N:144) | 0,0000 (L:1\|N:143) | 0,0000 (L:4\|N:140) | 0,0000 (B:1\|N:144) | 0,0018 (B:9\|N:144) | 0,0015 (B:2\|N:144) |
| IN | 0,0031 (L:3\|N:141) | 0,0009 (L:4\|N:140) | 0,0000 (L:4\|N:140) | 0,0000 (B:9\|N:144) | 0,0000 (B:10\|N:144) | 0,0000 (B:10\|N:144) |
| IR | 0,0067 (L:4\|N:140) | 0,0011 (L:4\|N:140) | 0,1113 (L:2\|N:142) | 0,0000 (B:7\|N:144) | 0,0000 (B:3\|N:144) | 0,0652 (B:8\|N:144) |
| IT | 0,0061 (L:8\|N:136) | 0,0001 (L:9\|N:135) | 0,0024 (L:0\|N:144) | 0,0673 (B:9\|N:144) | 0,0881 (B:10\|N:144) | 0,0027 (B:6\|N:144) |
| JP | 0,0001 (L:1\|N:143) | 0,0086 (L:2\|N:142) | 0,6762 (L:0\|N:144) | 0,0000 (B:9\|N:144) | 0,0000 (B:8\|N:144) | 0,3394 (B:6\|N:144) |
| KO | 0,0001 (L:3\|N:141) | 0,0149 (L:2\|N:142) | 0,1671 (L:2\|N:142) | 0,0000 (B:8\|N:144) | 0,0000 (B:9\|N:144) | 0,0440 (B:5\|N:144) |
| MX | 0,0415 (L:0\|N:144) | 0,3655 (L:9\|N:135) | 0,0414 (L:4\|N:140) | 0,0672 (B:7\|N:144) | 0,0901 (B:6\|N:144) | 0,0567 (B:6\|N:144) |
| MY | 0,0000 (L:0\|N:144) | 0,0027 (L:1\|N:143) | 0,0006 (L:0\|N:124) | 0,0000 (B:4\|N:144) | 0,0000 (B:8\|N:144) | 0,001 (B:4\|N:124) |
| NL | 0,0274 (L:3\|N:141) | 0,0040 (L:4\|N:140) | 0,1580 (L:12\|N:132) | 0,0000 (B:10\|N:144) | 0,0000 (B:10\|N:144) | 0,0557 (B:5\|N:144) |
| NW | 0,0602 (L:3\|N:141) | 0,0000 (L:3\|N:141) | 0,2180 (L:0\|N:144) | 0,0000 (B:9\|N:144) | 0,0000 (B:10\|N:144) | 0,0848 (B:5\|N:144) |
| OE | 0,0059 (L:4\|N:140) | 0,0101 (L:3\|N:141) | 0,0842 (L:0\|N:144) | 0,0000 (B:9\|N:144) | 0,0000 (B:9\|N:144) | 0,1008 (B:1\|N:144) |
| PH | 0,0002 (L:2\|N:142) | 0,0005 (L:2\|N:142) | 0,0001 (L:4\|N:140) | 0,0000 (B:7\|N:144) | 0,0000 (B:7\|N:144) | 0,0000 (B:9\|N:144) |
| PO | 0,0964 (L:9\|N:108) | 0,2312 (L:6\|N:112) | 0,0066 (L:2\|N:106) | 0,0001 (B:3\|N:117) | 0,0076 (B:3\|N:118) | 0,0768 (B:6\|N:108) |
| PT | 0,5028 (L:7\|N:137) | 0,3723 (L:7\|N:137) | 0,1767 (L:1\|N:143) | 0,0000 (B:10\|N:144) | 0,0005 (B:10\|N:144) | 0,0000 (B:9\|N:144) |
| RM | 0,0000 (L:0\|N:101) | 0,0000 (L:0\|N:102) | 0,0001 (L:4\|N:108) | 0,0000 (B:8\|N:101) | 0,0000 (B:8\|N:102) | 0,0022 (B:7\|N:112) |
| RS | 0,1050 (L:2\|N:97) | 0,1158 (L:1\|N:99) | 0,0004 (L:4\|N:100) | 0,0002 (B:3\|N:99) | 0,0410 (B:2\|N:100) | 0,0006 (B:7\|N:104) |
| SA | 0,0364 (L:2\|N:142) | 0,2099 (L:2\|N:142) | 0,0000 (L:5\|N:139) | 0,0000 (B:7\|N:144) | 0,0007 (B:9\|N:144) | 0,0000 (B:10\|N:144) |
| SD | 0,1198 (L:3\|N:141) | 0,0463 (L:3\|N:141) | 0,1913 (L:1\|N:143) | 0,0000 (B:9\|N:144) | 0,0002 (B:9\|N:144) | 0,1991 (B:7\|N:144) |
| SP | 0,0000 (L:0\|N:144) | 0,0035 (L:3\|N:141) | 0,0000 (L:1\|N:143) | 0,0000 (B:4\|N:144) | 0,0000 (B:3\|N:144) | 0,0159 (B:10\|N:144) |
| SW | 0,0496 (L:4\|N:140) | 0,1493 (L:3\|N:141) | 0,0000 (L:1\|N:143) | 0,0000 (B:10\|N:144) | 0,0000 (B:9\|N:144) | 0,0228 (B:7\|N:144) |
| TH | 0,0000 (L:0\|N:144) | 0,0003 (L:1\|N:143) | 0,0000 (L:4\|N:140) | 0,0000 (B:7\|N:144) | 0,0000 (B:6\|N:144) | 0,0002 (B:1\|N:144) |
| TK | 0,5735 (L:3\|N:141) | 0,6090 (L:2\|N:142) | 0,0000 (L:8\|N:136) | 0,0000 (B:9\|N:144) | 0,0099 (B:7\|N:144) | 0,0000 (B:10\|N:144) |
| TW | 0,0000 (L:3\|N:141) | 0,0042 (L:6\|N:138) | 0,1369 (L:5\|N:139) | 0,0000 (B:7\|N:144) | 0,0000 (B:6\|N:144) | 0,1121 (B:9\|N:144) |
| UK | 0,0117 (L:4\|N:140) | 0,0089 (L:4\|N:140) | 0,0413 (L:2\|N:142) | 0,0000 (B:10\|N:144) | 0,0000 (B:9\|N:144) | 0,0064 (B:8\|N:144) |
| US | 0,0002 (L:2\|N:142) | 0,0000 (L:4\|N:140) | 0,0295 (L:5\|N:139) | 0,0000 (B:7\|N:144) | 0,0000 (B:3\|N:144) | 0,0022 (B:6\|N:144) |
| VE | 0,0074 (L:0\|N:143) | 0,4144 (L:0\|N:111) | 0,0000 (L:4\|N:140) | 0,0095 (B:8\|N:143) | 0,2463 (B:2\|N:111) | 0,0000 (B:9\|N:144) |

**Notes:** The numbers in the table are rejection probabilities of the null hypotheses of ADF and PP tests including intercept. Probabilities below 0,10 denote a rejection of these null hypotheses, thus, confirm stationarity of the CPI (inflation), EI (expected inflation), and U_U' (unemployment gap) series of related countries. The lag and observation parameters are presented in parentheses where "L", "B", and "N" denote lag length, Newey-West bandwidth using Bartlett kernel, and observation number respectively. The lag length is determined by Schwarz Information Criterion (SIC) under a maximum lag length specification of 13.